\documentclass[12pt]{article}

\begin{document}
\begin{titlepage}
\title{\vskip -70pt
\begin{flushright}
{\normalsize \ DAMTP-97-90}\\
\end{flushright}
\vskip 20pt
{\bf Volume of Vortex Moduli Spaces }
}
\vspace{1cm}
\author{{N. S. Manton}\thanks{e-mail address: N.S.Manton@damtp.cam.ac.uk}
\hspace{.2cm}  and
\hspace{.2cm}
 {S. M. Nasir}\thanks{e-mail address: S.M.Nasir@damtp.cam.ac.uk}\\
{\sl Department of Applied Mathematics and Theoretical Physics}\\
{\sl University of Cambridge} \\
{\sl Silver Street, Cambridge CB3 9EW, England}}
\date{July, 1998}
\maketitle
\thispagestyle{empty}
\vspace{1cm}
\begin{abstract}
\noindent
A gas of $N$ Bogomol'nyi vortices in the Abelian
Higgs model is studied on a
compact Riemann surface of genus $g$ and area $A$.
The volume of the moduli space is computed and found to depend on
$N,\; g$ and $A$, but not on other details of the shape of the surface.
The volume is then used to find the
thermodynamic partition function and  it is shown that 
the thermodynamical properties of such a gas do not depend on the genus of
the Riemann surface.

\end{abstract}
\end{titlepage}

\section{Introduction}

Solitons are interesting objects to study and it is particularly
interesting to study their dynamics. The moduli space
approximation \cite{Man1} gives an effective description of the 
dynamics of solitons at low energy when most of the degrees of
freedom are frozen. The moduli space
approximation works as follows: static multi-solitons are parametrized by
the moduli space $-$ the minima of the energy functional. At low energy,
the  actual field dynamics can be taken to be close to the moduli
space, i.e. near the
bottom of the valley of the energy functional. The dynamics projected
onto the moduli space is then
the geodesic motion on the moduli space \cite{Stu}. 
For monopoles $-$ solitons in three dimensions $-$ the
moduli space approximation has given important insight into 
the
scattering and the bound states of the monopoles. It has found important
application in proving various duality conjectures in 
supersymmetric field theories and in string theory \cite{Sen}.

On a plane, for Bogomol'nyi
vortices in the Abelian Higgs model \cite{Bog} $-$ solitons in two
dimensions $-$ one
can similarly describe
their scattering \cite{Sam}. For vortices on a compact Riemann
surface, $M$, of genus $g$,
one can do
more $-$ study their statistical mechanics \cite{Man2}. Since the
potential energy between the
vortices is constant, in the moduli space approximation
evaluation of the partition function of a gas of
vortices effectively reduces to the computation of
the volume of the moduli space.
As the moduli space is K\"{a}hler, in
order to find the volume, one needs to know the K\"{a}hler form or
more precisely, its cohomology class. For the genus $g=0,1$ cases,
the K\"{a}hler forms have been computed in \cite{Man2, Sha},
respectively. For the sphere $(g=0)$ the $N$-vortex moduli space is
the complex projective
space $CP_{N}$.
In this case symmetry arguments are enough to find the K\"{a}hler
form. On the other hand, for the torus $(g=1)$, the K\"{a}hler form
is found
by exploiting the fibre bundle structure of the $N$-vortex moduli space.
For genus $g\geq 1$ and $N\geq 2g-1$, the $N$-vortex moduli space has a bundle
structure, where the base is the Jacobian, $J$, of the Riemann surface, a
torus of real dimension $2g$, and the fibre is $CP_{N-g}$. For
$N\leq g$, the $N$-vortex moduli space is homeomorphic to a
$2N$-dimensional analytic subvariety of the Jacobian. 
It would be interesting to find a general
formula for the
K\"{a}hler form and, hence, the volume of the moduli space for $N$
vortices on an
arbitrary Riemann surface, $M$, of genus $g$. Here, we will obtain
such a formula.

In the next section we will
see that the K\"{a}hler form is the sum of two parts: one is related
to the
K\"{a}hler form of $M$, the other is determined by the vortex interactions.
The cohomology classes of both of these can be determined. 
This then enables the required formula for the volume of the moduli
space to be computed. It depends on $N,\, g$ and the area of $M$.
The various thermodynamical quantities for a
gas of vortices can be deduced from this. 
It is found that the statistical mechanics of such a gas is
independent of the genus of $M$.
This is
expected on physical grounds. 

This paper is organized as follows. In sect.2, we briefly describe
the Bogomol'nyi vortices and the K\"{a}hler form on the moduli
space. In sect.3,  we present the cohomological formula for the volume. Then we
compare it with the 
previously computed cases for vortices on the sphere and the torus. This
serves as a check of the volume formula. Finally, in
sect.4, we compute the various thermodynamical quantities.

\section{Vortices and the K\"{a}hler form on the moduli space}
\setcounter{equation}{0}

\noindent
{\sl (i) Bogomol'nyi vortices}

Bogomol'nyi vortices are static, topologically stable, finite
energy solution of the critically coupled Abelian Higgs model in 2+1
dimensions \cite{Jaf}. We consider vortices on the space-time ${\bf{R}}\times
M$, where $M$ is a compact Riemann surface of genus $g$, and
${\bf{R}}$ parametrizes ordinary time $x_{0}$. The metric on
${\bf{R}}\times M$
can be taken to be of the form (locally)
\begin{equation}
ds^{2}=dx_{0}^{2}-\Omega (x_{1},x_{2})(dx_{1}^{2}+dx_{2}^{2}) 
\end{equation}
where $\Omega$ is positive.
The
Lagrangian density of the model is
\begin{equation}
{\mathcal{L}}=     -\frac{1}{4}
F_{\mu\nu}F^{\mu\nu} + \frac{1}{2}D_{\mu}\phi\overline{D^{\mu}\phi}
-\frac{1}{8}(|\phi|^{2}-1)^{2}
\end{equation}
where $\phi$ is a complex Higgs scalar field, $A_{\mu}$ is a $U(1)$
gauge potential,
$D_{\mu}=\partial_{\mu}-iA_{\mu}$ and $F_{\mu\nu}=\partial_{\mu}
A_{\nu}-\partial_{\nu}A_{\mu}\,\,\,(\mu,\nu=0,1,2)$. The units are chosen
such that both the gauge field coupling constant
and the mass of the Higgs field are unity.

Working in the gauge $A_{0}=0$, the Lagrangian is $L=T-V$ where
\begin{equation}\label{KE}
T=\frac{1}{2}\int_{M}
d^{2}x \,
(\dot{A}_{1}\dot{A}_{1}+\dot{A}_{2}\dot{A}_{2}+
 \Omega \dot{\phi}\dot{{\bar{\phi}}})
\end{equation}
\begin{equation}\label{PE}
V=\frac{1}{2}\int_{M} d^{2}x \left(\Omega^{-1}F_{12}^{2}+
D_{1}\phi\overline{D_{1}\phi}+D_{2}\phi\overline{D_{2}\phi}
+\frac{\Omega}{4}(|\phi|^{2}-1)^{2}\right)
\end{equation}
are respectively the kinetic  and the potential energies. Further, we need to
impose Gauss's law. This arises from the
equation of motion of $A_{0}$, as the following constraint,
\begin{equation}
\partial_{1}\dot{A}_{1}+\partial_{2}\dot{A}_{2}-
\Omega \, Im(\dot{\phi}{\bar{\phi}})=0 .
\end{equation}

\noindent
In the static case the total energy,
$E=V$, can be reexpressed as \cite{Bog}
\begin{equation}\label{Ene}
E = {\displaystyle{\frac{1}{2}}}\int_{M}\! d^{2}x\, \left( \, (D_{1}+
iD_{2})\phi\,\overline{(D_{1}+
iD_{2})\phi}+\Omega ^{-1} \{F_{12}+ {\displaystyle{\frac{\Omega}{2}}}
(|\phi|^{2}-1)\}^{2}
+ F_{12} \right) .
\end{equation}
Here we have omitted a total derivative term, which vanishes as $M$ has
no boundary. 
Bogomol'nyi vortices, which minimize the above energy integral,
satisfy the first order Bogomol'nyi equations 
\begin{equation}\label{Bog1}
(D_{1}+ iD_{2})\phi=0
\end{equation}
\begin{equation}\label{Bog2}
F_{12} + \frac{\Omega}{2}(|\phi|^{2}-1)=0.
\end{equation} 
The solutions are classified into topologically stable sectors
determined by the first Chern number \cite{Jaf, Tau}
\begin{equation}
\frac{1}{2\pi} \int_{M} d^{2}x F_{12} =N
\end{equation}
where $N$ is an integer. 
Note that, in general,
there is an obstruction
to the
existence of $N$-vortex solutions on a compact surface. This is
seen by integrating (2.8) over $M$. Since $\Omega |\phi |^{2}$ is
non-negative, we deduce the bound, first obtained by Bradlow
\cite{Bra},
\begin{equation}
4\pi N \leq A
\end{equation}
where $A$ is the area of $M$.
Assuming that this is satisfied, the solutions with the first Chern
number $N$ are uniquely determined
by specifying $N$ zeros of the Higgs field [2, 14]. Thus, $N$ can also be
interpreted as the vortex number. Since vortices are
indistinguishable, the vortex moduli
space, $M_{N}$, is diffeomorphic to the symmetric product
$(M)^{N}/S_{N}$ where $S_{N}$ is the
permutation group of $N$ elements. It should be noted that $M_{N}$ is
a smooth manifold. In the sector with vortex
number $N$, the potential energy of the vortices is $E=\pi N$.

It is possible to eliminate the gauge potentials from eqn.(2.8), by
solving (2.7), thereby obtaining an equation for $|\phi |^{2}$
\begin{equation}
\nabla^{2} \log |\phi |^{2}-\Omega (|\phi |^{2}-1)=4\pi \sum_{i=1}^{N}
\delta^{2} ({\bf{x}}-{\bf{x}}_{i})
\end{equation}
where ${\bf{x}}_{i}$ denotes the position of the zero of the Higgs field
associated with the $i$-th vortex and
$\nabla^{2}=\partial_{1}^{2}+\partial_{2}^{2}$.

The kinetic energy (2.3) induces a natural Riemannian metric on the
moduli space \cite{Man1}. Let $q_{\alpha}$ and $g_{\alpha
\beta}({\bf{q}})dq^{\alpha}dq^{\beta}
$, where $(\alpha , \beta =1,\cdots ,2N)$ denote 
real coordinates and
the metric on $M_{N}$. Then, in the moduli space approximation
for vortex motion the Lagrangian can be
written as
\begin{equation}
L=\frac{\pi}{2}g_{\alpha \beta
}({\bf{q}})\dot{q}^{\alpha}\dot{q}^{\beta} -N\pi
\end{equation}
where $\pi$ is the mass of a single vortex. Below, we shall use the
analogue of this expression using complex coordinates for the vortex
positions. Although we cannot
determine $g_{\alpha \beta}$ explicitly, we shall show that it is possible to
compute the total volume of $M_{N}$.

\noindent
{\sl (ii) The K\"{a}hler form on the moduli space}

Samols \cite{Sam} has obtained an expression for the metric $g_{\alpha
\beta}$
and the associated K\"{a}hler form on
the $N$-vortex moduli space by analyzing data around the $N$ zeros of the Higgs
field assuming these are distinct. Detailed computation shows that the
metric has a smooth extension to the complete moduli space, where
vortices may coincide. Let $z$ be a local complex
coordinate on $M$. 
Let the vortex positions be 
$\{ z_{i}=x_{1i}+ix_{2i}:i=1,\cdots ,N\}$. Since $z_{i}$ is a simple
zero of the Higgs field,
$\log |\phi |^{2}$ has the following series expansion obtained on
using (2.11),
\begin{eqnarray}\label{hex}
\log |\phi |^{2} =\log|z-z_{i}|^{2} + a_{i} \! \! \! & + & \! \! \!
\frac{1}{2}b_{i}(z-z_{i})+
\frac{1}{2}\bar{b_{i}}(\bar{z}-\bar{z_{i}}) + c_{i}(z-z_{i})^{2} \nonumber \\
                 & - & \! \! \! \frac{\Omega (z_{i})}{4}(z-z_{i})
(\bar{z}-\bar{z_{i}})+\bar{c_{i}}(\bar{z}-\bar{z}_{i})^{2}+\cdots 
.
\end{eqnarray}
From the expression for the kinetic energy eqn.(2.3), Samols shows,
after some integrations,
that the metric is
\begin{equation}\label{Met}
ds^{2}=\sum_{i,j=1}^{N}\left(\Omega (z_{i})\delta_{ij}+
2\frac{\partial
b_{i}}{\partial {\bar{z}}_{j}}\right)dz_{i}d\bar{z_{j}}.
\end{equation}
Only the coefficients of the linear
terms in (2.13) contribute to this formula. Notice that $b_{i}$ is
a function of the positions of all $N$ vortices.

The reality property of the kinetic energy implies that
\begin{equation}
\frac{\partial \bar{b}_{i}}{\partial z_{j}}=\frac{\partial
b_{j}}{\partial \bar{z}_{i}}
\end{equation}
and from this follows the Hermiticity of the metric (2.14). One
can then define
the associated K\"{a}hler form as
\begin{equation}
\omega=\frac{i}{2}\sum_{i,j=1}^{N}\left(\Omega (z_{i})\, \delta_{ij}+
2\frac{\partial
b_{i}}{\partial {\bar{z}}_{j}}\right)dz_{i}\wedge d\bar{z_{j}}.
\end{equation}
Using (2.15) one can show that $\omega$ is a closed (1,1) form. 
The volume of the moduli
space is
\begin{equation}
{\rm{Vol}}_{N}=\frac{1}{N!}{\int}_{M_{N}} \omega^{N} .
\end{equation}
The K\"{a}hler form $\omega$ can be  divided into two parts
$\omega=\omega_{1}+\omega_{2}$, where
\begin{equation}
\omega_{1}=\frac{i}{2}\sum_{i=1}^{N}\Omega (z_{i})dz_{i}\wedge d\bar{z_{i}}
\end{equation}
is just $N$ copies of
the area form induced from $M  $ and 
\begin{equation}
\omega_{2}=i \sum_{i,j=1}^{N}\frac{\partial
b_{i}}{\partial {\bar{z}}_{j}}dz_{i}\wedge d\bar{z_{j}}
\end{equation}
contains information about the relative
vortex positions. Our aim is to understand the topological nature of
$\omega_{2}$ and its effect on ${\rm{Vol}}_{N}$. If $\omega_{2}$ is
ignored, ${\rm{Vol}}_{N}$ would simply be $A^{N}/N!$. Notice that to
obtain this result we have chosen a specific normalization of
$\omega$ dictated by physics. In fact we can choose any normalization
by multiplying the Lagrangian by an overall constant.

Notice that one can write $\omega_{2}=-i{\bar{\partial}} B$ where $B$ is
a one-form of degree (1,0)
\begin{equation}
B=\sum_{i=1}^{N}b_{i}(z_{1},z_{2},\cdots
,z_{N},{\bar{z}}_{1},{\bar{z}}_{2},\cdots ,{\bar{z}}_{N})dz_{i}.
\end{equation}
Since $z_{i}$ are natural coordinates on the Cartesian
product $(M)^{N}$, not on the moduli space $M_{N}$, 
the symmetry of the one-form $B$ is not manifest in the above equation.
However,
the indistinguishability of vortices implies that 
\begin{equation}
b_{i}(\cdots ,z_{i}, \cdots ,z_{j}, \cdots)=b_{j}(\cdots ,z_{j},
\cdots, z_{i}, \cdots )\; .
\end{equation}
Thus, the 
one-form $B$ is symmetric and hence, defined on $M_{N}$. 

Before proceeding further we would like to point out that the one-form
$B$ has poles
whenever $z_{i}=z_{j}$ for $i\neq j$.
To see this let us consider the function $\psi$
defined in a coordinate patch as follows
\begin{equation}
\psi =\log |\phi|^{2} -\sum_{i=1}^{N}\log |z-z_{i}|^{2}.
\end{equation}
Notice that $\psi$ is
a smooth function, since the singularities of $\log |\phi|^{2}$ at the
zeros of the Higgs field have been cancelled by the term 
$\sum_{i=1}^{N}\log |z-z_{i}|^{2}$. Then, as $z_{j}$ approaches
$z_{i}$, one can see that
\begin{equation}
b_{i}=\frac{2}{z_{i}-z_{j}} +{\rm{smooth\; \; part}},
\end{equation}
hence, $B$ has poles.
It is useful to note
that the residues of $B$ are
integers. This fact will be important later.

One simple way to uncover the topological significance of
$B$ is to determine its transformation
properties 
under change of coordinates. Let us assume that
$M  $ is covered in such a way
that all $N$ vortices lie in one coordinate patch $U$ whose local
coordinate is denoted by $z$. The $i$-th vortex in this patch has the
coordinate $z_{i}$. 
Under a holomorphic coordinate transformation, $U$ goes into another
coordinate patch $U'$.
In terms of the local coordinate $z\rightarrow z'=\zeta (z)$;
and, also $z_{i}\rightarrow z'_{i}=\zeta (z_{i})$. In the
transformed coordinate, the expansion of $\log |\phi|^{2}$ in (2.13) reads
\begin{eqnarray}
\log |\phi |^{2} = \log|z'-z'_{i}|^{2}  +  a'_{i} \! \! \! & + & \! \!
\!
\frac{1}{2}b'_{i}(z'-z'_{i})+
\frac{1}{2}\bar{b'_{i}}(\bar{z'}-\bar{z'_{i}})+
c'_{i}(z'-z'_{i})^{2} \nonumber \\
                    & - & \! \! \! \frac{\Omega (z'_{i})}{4}(z'-z'_{i})
(\bar{z'}-\bar{z'_{i}})+\bar{c'_{i}}(\bar{z'}-\bar{z'}_{i})^{2}+\cdots
.
\end{eqnarray}
Here, 
$b'_{i}=b'_{i}(z'_{1},\cdots ,z'_{i},\cdots ,z'_{N})$, writing out
the coordinate dependence explicitly. Remember that $|\phi |^{2}$ is a
globally well defined function on $M$. Thus, on the
overlap region $U\cap U'$,
by comparing the coefficients of $(z-z_{i})$
on the right hand sides of eqns. (2.13) and (2.24) one finds
\begin{equation}
b_{i}=b'_{i}\frac{\partial \zeta_{i}}{\partial z_{i}}+\frac{\partial
z_{i}}{\partial \zeta_{i}}\frac{\partial ^{2}\zeta_{i}}{\partial z_{i}^{2}}
\end{equation}
where $\zeta_{i}=\zeta(z_{i})$. Notice the striking similarity with the
corresponding transformation of the Levi-Civita connection of $M  $
\begin{equation}
\Gamma ^{z}_{zz}=\Gamma^{z'}_{z'z'}\frac{\partial z'}{\partial z}+
\frac{\partial
z}{\partial z'}\frac{\partial ^{2}z'}{\partial z^{2}}.
\end{equation}
This heralds the topological nature of $B$. By looking at equations (2.25)
and (2.26), 
one concludes
that $B$ differs from the complex connection one-form on the
co-tangent bundle of
$M_{N}$ by a globally defined one-form. Generically, this one-form is
not smooth as it contains
poles. If the poles were absent then
$\omega_{2}=-i{\bar{\partial}}B$
would have been cohomologous to the 
complex Ricci curvature two-form of the Levi-Civita connection on the
co-tangent bundle of $M_{N}$. This means that $
\omega_{2}/2\pi$ would have been
cohomologous to the 
first
Chern class 
of the
co-tangent bundle.

In what follows we will need to evaluate the integrals of
$\omega_{2}$ restricted to some special complex one-dimensional submanifolds. 
The integrals, as we will see shortly, receive two  
contributions: one is from the residues of $B$, and the other is due
to the fact that $B$, restricted to these submanifolds, is related
to the complex Levi-Civita connection.

First, we consider the submanifold of $N$ coincident vortices.
The solutions with $N$ coincident vortices are parametrized by
a complex one-dimensional
submanifold, $M_{co}$ of the moduli space $M_{N}$. $M_{co}$ is
diffeomorphic to $M$ and lies inside the Jacobian $J$. Let $Z$ be the
position of
the coincident vortices. $|\phi |^{2}$ now satisfies the equation 
\begin{equation}
4\frac{\partial^{2} \log |\phi |^{2}}{\partial z \partial {\bar{z}}}
-\Omega (|\phi |^{2}-1)=4\pi N
\delta^{2} (z-Z)
\end{equation}
and $\log |\phi |^{2}$ has the following
series expansion around $Z$
\begin{equation}
\log |\phi |^{2}=N\log |z-Z|^{2} +a
+\frac{1}{2}b(z-Z)+\frac{1}{2}{\bar{b}}({\bar{z}}-{\bar{Z}})+ \cdots \; \;
.
\end{equation}
Then, the one-form $B$, restricted to $M_{co}$, simplifies to
\begin{equation}
B=b(Z,{\bar{Z}})dZ.
\end{equation}
By a similar analysis as in
(2.25), one can determine the transformation properties of $b$
under a holomorphic coordinate transformation
$z\rightarrow \xi (z)$. One obtains
\begin{equation}
\frac{b}{N}=\frac{b'}{N}\frac{\partial \xi (Z)}{\partial Z}+\frac{\partial
Z}{\partial \xi (Z)}\frac{\partial ^{2}\xi (Z)}{\partial Z^{2}}.
\end{equation}
We remark that for $N$ coincident vortices $B$ does not contain any pole.
By comparing (2.30) with (2.26), one finds that $B/N$ restricted
to the submanifold $M_{co}$ differs from the 
complex Levi-Civita connection one-form of $M_{co}
$ by a smooth, globally defined one-form. Thus, $\omega_{2} /N$
restricted to $M_{co}$
is cohomologous to the complex Ricci curvature
two-form of the co-tangent
bundle of $M_{co}$. Now,
the volume of $M_{co}$ is
\begin{equation}
V_{co}=\int_{M_{co}}\omega =\frac{iN}{2}\int _{M  }\left(\Omega +
2\frac{\partial b}{\partial
\bar{Z}}\right) dZ\wedge d\bar{Z} =N\left( A-4\pi N(1-g)\right)
\end{equation}
where use has been made of the Gauss-Bonnet formula for the integral
of the curvature of the Levi-Civita connection, $\Gamma^{Z}_{ZZ}$, on
$M$
\begin{equation}
\frac{-i}{2\pi }\int_{M}
\left(\displaystyle{\frac{\partial \Gamma^{Z}_{ZZ}}{\partial
\bar{Z}}}\right)  dZ\wedge
d\bar{Z} =2(1-g)
\end{equation}
which implies
\begin{equation}
\frac{-i}{2\pi N}\int_{M}
\left(\displaystyle{\frac{\partial b}{\partial
\bar{Z}}}\right)  dZ\wedge
d\bar{Z} =2(1-g).
\end{equation}
Notice that ${\displaystyle{\frac{1}{2\pi}\int_{M_{co}}\omega_{2}=
2N^{2}(g-1)}}$.
The volume $V_{co}$ agrees with the volumes
previously computed for the sphere and the torus in [7, 12],
respectively. 

Secondly, let us consider two clusters of vortices with
$m$ and $(N-m)$ vortices, and let
$z_{1}$ and $z_{2}$ be their positions on $M$, respectively. The solutions
corresponding to such clusters are parametrized by a complex two-dimensional
submanifold, $M_{c},$ of the moduli space $M_{N}$.
We can do a similar analysis as in the above to compute the integral
of $\omega_{2}$ restricted to certain one-dimensional submanifolds
of $M_{c}$.
Restricted to $M_{c}, \; B$ can be written as
\begin{equation}
B=b_{1}dz_{1}+b_{2}dz_{2}.
\end{equation}
Notice that $b_{1}$
has a pole at $z_{1}=z_{2}$
and from the generalization of (2.23) to a pair of vortex clusters,
one finds that ${\rm{Res}}(b_{1})=2(N-m)$.
Following (2.25) one can determine the transformation properties of $b_{1}$
and $b_{2}$
under holomorphic coordinate changes $z_{1}\rightarrow z'_{1}
$ and $z_{2}\rightarrow z'_{2}$. These are
\begin{equation}
b_{1}(z_{1},z_{2}) =b'_{1}(z'_{1},z'_{2})\frac{\partial
z'_{1}}{\partial z_{1}} +m\frac{\partial
z_{1}}{\partial z'_{1}}\frac{\partial ^{2}z'_{1}}{\partial z_{1}^{2}} 
\end{equation}
\begin{equation}
b_{2}(z_{1},z_{2}) =b'_{2}(z'_{1},z'_{2})\frac{\partial
z'_{2}}{\partial z_{2}} +(N-m)\frac{\partial
z_{2}}{\partial z'_{2}}\frac{\partial ^{2}z'_{2}}{\partial z_{2}^{2}}.
\end{equation}
We will be particularly interested in the case when the second cluster
does not move, i.e. when $z_{2}$ is a constant. The vortex motion is
then restricted to a one-dimensional submanifold, ${\tilde{M}} $, of
$M_{c}$. ${\tilde{M}} $ is diffeomorphic to $M$.

Now, comparing (2.35) with (2.26) one sees that
$B/m$, restricted to $\tilde{M}$, differs from
the complex Levi-Civita connection
one-form of ${\tilde{M}} $ by a one-form which contains a pole at
$z_{1}=z_{2}$.  For the volume of $\tilde{M}$ one can write 
\begin{equation}
{\tilde{V}}=\int_{{\tilde{M}} }\omega=I_{r}+I.
\end{equation}
Here, $I_{r}$ is the
contribution coming from the residues and $I$ contains the
rest of the contribution.
Similarly as in the derivation of (2.31) we find
\begin{equation}
I=m\left( A-4\pi m(1-g)\right)
\end{equation}
and the residue contribution is
\begin{equation}
I_{r}=-2\pi m\, {\rm{Res}}(b_{1})=-4\pi m(N-m).
\end{equation}
Thus, the total volume of ${\tilde{M}} $ is
\begin{equation} 
{\tilde{V}}=m(A-4\pi N+4\pi mg).
\end{equation}
As a consistency check, if $m=N$ then we have one cluster of $N$
coincident vortices. 
In this case we get back (2.31) by simply
putting $m=N$ in the above formula.

We remark that
${\displaystyle{\frac{1}{2\pi}\int_{{\tilde{M}} }\omega_{2}=2m(mg-N)}}$.
Naturally, one would expect that the (1,1) form $\omega_{2}$ belongs
to $H^{2}(M_{N},{\bf{R}})$, since this is a part of the K\"{a}hler form
of $M_{N}$. However, because of the relationship between $B$ and the
complex Levi-Civita connection one-form,
combined with the fact that the residues of $B$ are integers, one
sees that the integral of $\omega_{2}/2\pi$ over any complex
one-dimensional submanifold is
an integer. This means that 
$\omega_{2}/2\pi$ actually belongs to $H^{2}(M_{N},{\bf{Z}})$. This
information will be used in the next section
in obtaining a cohomological formula for $\omega_{2}$.

\section{Cohomology and the volume of the moduli space}
\setcounter{equation}{0}

\noindent
{\sl (i) Cohomology ring of the symmetric products of a Riemann surface}

Here, we quote several theorems without proof which will be used
later. This also serves to fix the notation. The main reference 
is \cite{Mac}.

We have $H^{0}(M
,{\bf{Z}})={\bf{Z}},\;H^{1}(M,{\bf{Z}})={\bf{Z}}^{2g}$ and $ H^{2}(M  
,{\bf{Z}})={\bf{Z}}$. Let $\alpha_{i},\; i=1,\cdots ,2g$ be the generators of
$H^{1}(M,{\bf{Z}})$ and $\beta $ be the generator of $H^{2}(M  
,{\bf{Z}})$. It is useful to note that $\beta$ is a normalized area
form (i.e. its integral over $M$ is unity) of type (1,1).
The ring structure of $H^{\ast}(M,{\bf{Z}})$ can be described
as follows
\begin{equation}
\alpha_{i} \alpha_{j}=0, \;\;i\neq j\pm g\;\;
;\alpha_{i}\alpha_{i+g}=-\alpha_{i+g}\alpha_{i}=\beta , \; \;1\leq i
\leq g.
\end{equation}
Here, juxtaposition means cup product. Let
\begin{equation}
\begin{array}{lll}
\alpha_{ik} & = & 1\otimes \cdots \otimes 1\otimes \alpha_{i} \otimes
1 \otimes \cdots
\otimes 1\in H^{1}((M)^{N},{\bf{Z}})   \\
\beta_{k}  & = & 1\otimes \cdots \otimes 1\otimes \beta \otimes 1\otimes
\cdots \otimes 1\in H^{2}((M)^{N},{\bf{Z}}),
\end{array}
\end{equation}
the $\alpha_{i}$ and $\beta$ being in the $k$-th place. Then,
$H^{\ast}((M)^{N},{\bf{Z}})$ is generated by the $\alpha_{ik}$ and the
$\beta_{k}\; \;(1\leq i \leq 2g, 1\leq k \leq N)$ with
 the following
relations being satisfied
\begin{equation}
\begin{array}{lll}
\alpha_{ik}\alpha_{jk}    & = & 0,\; \; i\neq j\pm g  \\
\alpha_{ik}\alpha_{i+g,k} & = & -\alpha_{i+g,k}\alpha_{ik}=\beta_{k},\; \;
1\leq i \leq g \\
\alpha_{ik}\alpha_{jl}    & = & -\alpha_{jl}\alpha_{ik},\; k\neq l.
\end{array}
\end{equation}

\noindent
Now, define the following symmetric linear combinations
\begin{equation}
\begin{array}{lll}
\xi_{i} & = & \alpha_{i1}+\cdots +\alpha_{iN},\; \; 1\leq i\leq 2g \\
\eta    & = & \beta_{1}+\cdots +\beta_{N}.
\end{array}
\end{equation}
Further, define $\xi'_{i}=\xi_{i+g}\; \; (1\leq i \leq g)$ and
$\sigma_{i}=\xi_{i}\xi'_{i}$. Then we have the following result
\cite{Mac}

\vspace{.5cm}
\noindent
{\bf{Theorem 1}} {\sl Let $M  $ be a compact connected Riemann
surface of genus g, $M_{N}$ its $N$-th symmetric product. Then, the
cohomology ring $H^{\ast}(M_{N},{\bf{Z}})$ is generated by elements
$\xi_{1},\cdots ,\xi_{g},\xi'_{1},\cdots ,\xi'_{g}$ of degree 1, and an
element $\eta$ of degree 2, subject to the following relations:

  (a) the $\xi$'s and $\xi'$'s anti-commute with each other and commute
  with $\eta$;

  (b) If $i_{1},\cdots ,i_{a},\;j_{1},\cdots ,j_{b},\;k_{1},\cdots
  ,k_{c}$ are distinct integers from 1 to $g$ inclusive, then 
\begin{equation}
\xi_{i_{1}}\cdots \xi_{i_{a}}\xi'_{j_{1}}\cdots
\xi'_{j_{b}}(\sigma_{k_{1}}-\eta)\cdots
(\sigma_{k_{c}}-\eta) \eta^{q}=0
\end{equation}
provided that $a+b+2c+q=N+1$. }

We will also need the following result on the cohomology of some
particular submanifolds of $M_{N}$. Let $\nu=(N_{1}\cdot p_{1}+\cdots
+N_{k}\cdot p_{k})$ be a partition of $N$ such that $
p_{1}>p_{2}>\cdots p_{k}>0$ and
$N=\sum p_{i}N_{i}$. Then there exists a mapping from
$\prod_{i=1}^{k}M_{N_{i}}$ onto a closed submanifold $\triangle (\nu)$
of $M_{N}$, where $\triangle (\nu)$ has $N_{1}$ clusters of $p_{1}$
coincident vortices, $N_{2}$ clusters of $p_{2}$ coincident vortices, etc.
This mapping is an isomorphism. For any submanifold $Y$, let us write $[Y]$ for
its cohomology class in $H^{\ast}(M_{N},{\bf{Z}})$. Then, one can show
that \cite{Mac}

\noindent
{\bf{Theorem 2}} {\sl $[\triangle (\nu)]$ is the coefficient of 
$\tau_{1}^{N_{1}}\cdots \tau_{k}^{N_{k}}$ in 
\begin{equation}
P^{\rho -g}\eta^{N-\rho -g}\prod_{i=1}^{g}(P\eta+Q(\eta-\sigma_{i}))
\end{equation}
where
\begin{equation}
\begin{array}{lll}
P    & = & p_{1}\tau_{1}+\cdots +p_{k}\tau_{k}, \\
Q    & = & (p_{1}^{2}-p_{1})\tau_{1}+\cdots +(p_{k}^{2}-p_{k})\tau_{k}, \\
\rho & = & N_{1}+\cdots +N_{k}.
\end{array}
\end{equation}
}

Now, if
$\delta_{s}=[\triangle(1\cdot s+(N-s) \cdot 1)], \;s>1,$ so that
$\delta_{s}$ is
the cohomology class of the submanifold of $M_{N}$ which consists of
those points which have at least $s$ vortices coinciding at one point,
then one can
show using Theorem 2 that
\begin{equation}
\delta_{s}=s(N+(g-1)(s-1))\eta^{s-1}-s(s-1)\eta^{s-2}( \sigma_{1}+
\cdots +\sigma_{g}).
\end{equation}
In terms of the above notation, the submanifold
$M_{co}$ for $N$ coincident vortices corresponds to $\triangle(1\cdot
N)$ and its
cohomology class is
\begin{equation}
\delta_{N}=N(N+(g-1)(N-1))\eta^{N-1}-N(N-1)\eta^{N-2}(\sigma_{1}+\cdots
+\sigma_{g}).
\end{equation}

Further, the total Chern class of the tangent bundle of $M_{N}$ is
$(1+\eta)^{N-2g+1}\prod_{i=1}^{g}(1+\eta -\sigma_{i})$. So,
the first Chern class of the tangent bundle is
\begin{equation}
c_{1}(TM_{N})=(N-g+1)\eta -(\sigma_{1}+\cdots +\sigma_{g}).
\end{equation}

\noindent
{\sl (ii) Cohomological formula for the K\"{a}hler form and the volume
of the moduli space} {\footnote{An attempt to obtain a cohomological
formula for the K\"{a}hler form was first made by P. Shah
\cite{Sha1}. His work has inspired us to look further into the
problem from a cohomological point of view. }}

An expression for the cohomology class of the two-form $\omega_{2}$ can
be obtained using the fact that
$\omega_{2}/2\pi$ is a (1,1) form
belonging to
$H^{2}(M_{N},{\bf{Z}})$. Let us determine the generators of
$H^{2}(M_{N},{\bf{Z}})$ which are of type (1,1).
One can see that $\eta$
is a generator of $H^{2}(M_{N},{\bf{Z}})$, and this is of type (1,1).
The other type (1,1) generator of
$H^{2}(M_{N},{\bf{Z}})$ comes from the pairing of the generators of
$H^{1}(M_{N},{\bf{Z}}) $. 
In 
Appendix (i) we show that it must be of
the form
\begin{equation}
D'(\sigma_{1}+\cdots +\sigma_{g})
\end{equation}
where $D'$
is a non-zero integer.
Thus, the general expression for
$\omega_{2}$ reads
\begin{equation}
\omega_{2} =2\pi C(g,N)\eta +2\pi D(g,N)(\sigma_{1}+\cdots +\sigma_{g})
\end{equation} 
where $C(g,N)$ and $D(g,N)$ are integers.

The coefficients $C(g,N)$ and $D(g,N)$ can be determined by computing
the volumes of the
submanifolds describing different types of coincident vortices
by cohomological means and
then comparing them with the same volumes previously computed 
in sect.2. The volume of $M_{co}\;-$ which describes the motion of $N$
coincident vortices $-$ is
\begin{equation}
V_{co}=\int _{M_{co}}
(\omega_{1}+\omega_{2})=\int_{M_{N}}(\omega_{1}+\omega_{2})
\wedge \delta_{N}.
\end{equation}
Using (3.9) and (3.5), one finds 
\begin{equation}
V_{co}=N\left( A+2\pi C(g,N)+2\pi NgD(g,N)\right)
\end{equation}
where we have used the fact that 
\begin{equation}
\omega_{1}=A\eta .
\end{equation}
Equating this with (2.31), we require
\begin{equation}
C(g,N)+NgD(g,N)=2 N(g-1).
\end{equation}
In Appendix (ii) we show that the volume of the submanifold ${\tilde{M}} \;-$ which
describes the motion of $m$ coincident vortices with the remaining
$(N-m)$ vortices coincident and held fixed at a general position $-$ is
\begin{equation}
{\tilde{V}}=m\left( A+2\pi C(g,N)+2\pi mgD(g,N)\right) .
\end{equation}
Comparing this with (2.40) gives
\begin{equation}
C(g,N)+mgD(g,N)=-2 N+2 mg.
\end{equation}
From (3.16) and (3.18), we find
\begin{equation}
C(g,N)=-2 N, \;\; \; D(g,N)=2 .
\end{equation} 
Thus, the K\"{a}hler form on $M_{N}$ is
\begin{equation}
\omega=\omega_{1}+\omega_{2}=(A-4\pi N)\eta +4\pi (\sigma_{1}+\cdots +
\sigma_{g}).
\end{equation}
Notice that 
\begin{equation}
\omega_{2}/2\pi =-2c_{1}(TM_{N}) +2(1-g)\eta 
\end{equation}
where use has been made of (3.10).
This shows that $\omega_{2}/2\pi$ is not just the
first Chern class of the
co-tangent bundle of $M_{N}$.

Now, putting all the ingredients together, and using (3.5),
one finally obtains the following
formula for the volume of
the moduli space 
\begin{equation}
{\rm{Vol}}_{N}=\int_{M_{N}}\frac{\omega^{N}}{N!}=(A-4\pi
N)^{N-g}\sum_{i=0}^{g}\left(\frac{(4\pi)^{i}(A-4\pi N)^{g-i}g!}{(N-i)!
(g-i)!i!}\right) .
\end{equation}
In the formula above $N\geq g$. An analogous formula can be written
for $N<g$. The sum now runs from $i=0$ to $i=N$, and the factors of
$A-4\pi N$ are combined to give $(A-4\pi N)^{N-i}$ in the sum.
Notice that the volume is just a function of the area of $M$,
its genus, and the number of vortices. It contains no
information about the shape of $M$.

\noindent
{\sl (iii) Examples: the volume of the moduli space for the sphere and
the torus}

For the sphere $(g=0)$, (3.22) gives
\begin{equation}
{\rm{Vol}}_{N}=\frac{(A-4\pi N)^{N}}{N!}.
\end{equation}
This is precisely the same as the formula obtained in \cite{Man2}. On
the other hand for the torus $(g=1)$ one gets
\begin{equation}
{\rm{Vol}}_{N}=\frac{A(A-4\pi N)^{N-1}}{N!}.
\end{equation}
Again this is the same as the
formula obtained in \cite{Sha}.
Shah conjectured in [11] that the volume of the moduli space for any Riemann
surface with genus $g>1$ is given by (3.24). We, however, find this
conjecture to be not true,
e.g. for a Riemann surface of genus $g=2$ and $N\geq 2$, the volume is
\begin{equation}
{\rm{Vol}}_{N}=\frac{(A^{2}-16\pi^{2}N)(A-4\pi N)^{N-2}}{N!}
\end{equation}
which is different from (3.24).

\section{Thermodynamics of the vortices}
\setcounter{equation}{0}

Following [7], the thermodynamics of $N$
vortices at temperature $T$ can be treated using the Gibbs
distribution. The partition
function is
\begin{equation}
{\mathcal{Z}}=\frac{1}{h^{2N}}\int _{M_{N}} [d{\bf{p}}] [d{\bf{q}}]
e^{-E({\bf{p}},{\bf{q}})/T}
\end{equation} where $h$ is Planck's constant, $p_{\alpha}$ are the
momenta conjugate to the
coordinates $q_{\alpha}$ and $E$ is the energy. After doing the
Gaussian momentum integrals, the partition function reduces to
\begin{equation}
{\mathcal{Z}}=(2\pi^{2}T/h^{2})^{N}\int_{M_{N}} [d{\bf{q}}]
({\rm{det}}g_{\alpha \beta })^{1/2}.
\end{equation}
The second factor in this partition function is just the volume,
${\rm{Vol}}_{N}$, 
of the moduli space $M_{N}$.

Using (4.2) and (3.22) one obtains the partition function for a gas of $N$
vortices on $M$
\begin{equation}
{\mathcal{Z}}=\frac{(A-4\pi N)^{N-g}}{N!}\left(\frac{2\pi
^{2}T}{h^{2}}\right)^{N}R(g,A,N)
\end{equation}
where 
\begin{equation}
R(g,A,N)=\sum_{i=0}^{g}\frac{(A-4\pi
N)^{g-i}(4\pi)^{i}g!N!}{(N-i)!(g-i)!i!}.
\end{equation}
To obtain the thermodynamic limit, we let $N\rightarrow
\infty$, assuming that the density of the gas of 
vortices is a fixed constant given by $N/A=n$. Now, a short
calculation shows that, at fixed $n$,
\begin{equation}
R(g,A,N)=A^{g}\left(1+O(1/N)\right).
\end{equation}
Using Stirling's formula for $N!,$ when $N$ is large, one obtains the
free energy $F=-T\log {\mathcal{Z}}$,
\begin{equation}
F\simeq -NT\left(\log \frac{2e\pi^{2}T}{h^{2}}-\log N+(1-\frac{g}{N})\log
(A-4\pi N)+ \frac{g}{N}\log A
+O(1/N) \right).
\end{equation}
The pressure $P=-\partial
F/\partial A$ is
\begin{equation}
P=\frac{NT}{A-4\pi N}.
\end{equation}
The entropy $S=-\partial F/\partial T$ is
\begin{equation}
S=N\left( \log \left( \frac{1-4\pi n}{n}\right)+\log 
\left(\frac{2e^{2}\pi^{2}T}{h^{2}}\right) \right).
\end{equation}
These are precisely the same formulae as obtained in [7, 12].
Notice that the genus $g$ appears nowhere in the formulae for
the thermodynamical quantities. Thus, the thermodynamics of a gas of
vortices is independent of
the topology of the space on which the vortices are moving.

\section{Conclusion}
\setcounter{equation}{0}
Central to our study of the thermodynamics of a gas of vortices on an
arbitrary Riemann surface is the
computation of the volume of the vortex moduli space. The dependence
of the volume on the area of the Riemann surface is quite noticeable. 
The area dependence disappears from the volume
whenever $A=4\pi N\,-$ Bradlow's limit. Then,
for $N\leq g$ the
volume of the
moduli space is $
{\rm{Vol}}_{N}=(4\pi )^{N}g!/[N!(g-N)!]$,
and for $N>g$ the volume is zero.
At $A=4\pi N$ the Higgs field
vanishes everywhere and
the problem of solving the Bogomol'nyi equations reduces to the problem of
solving for a constant magnetic field on the Riemann surface $M$. 
It can be shown that for $N=g$ the moduli space of this problem is related to
the space of
flat $U(1)$
connections on $M$. Time-varying flat connections have non-trivial
kinetic energy, and hence, following the argument of sect.2, there is
a metric on this moduli space. The volume of this moduli
space is a topological quantity.
It is of interest to
see that the volume of this moduli space is equal to 
${\rm{Vol}}_{g}$ at Bradlow's limit. This is shown
in Ref.\cite{Nas}. For $N>g$, it is also shown in \cite{Nas}, how
${\rm{Vol}}_{N}$ tends to zero as $A$ approaches $4\pi N$.

Moduli spaces play an important role in diverse areas of physics and
mathematics.
In general 
it is desirable
to know more about moduli spaces,
e.g. their volume (compact cases), metric etc.
Computation of the volume of a moduli space is not totally new.   
In
\cite{Witten}, with a remarkable use of the Verlinde formula \cite{Ver},
Witten computed
the volume of the moduli space of flat connections (for semi-simple
gauge groups) on an arbitrary
Riemann surface. In this case, however, the volume
is a purely topological quantity.
Thus, it is gratifying to see that in the case of the moduli space of
Bogomol'nyi vortices on a compact Riemann surface 
one can also explicitly compute the
volume. This is almost topological, but not exactly so, because
the volume depends on the area of the Riemann surface, not on its shape.

\appendix
\section*{Appendix}
\setcounter{section}{1}
\setcounter{equation}{0}

\noindent
{\sl (i) A note on a (1,1) form belonging to
$H^{2}(M_{N},{\bf{Z}})$}

Let $w_{\rho},\; (\rho=1,\cdots ,g)$ be a basis of holomorphic
one-forms on $M$ with the period matrix $\Lambda =(\lambda_{\rho
i}),\; (i=1,\cdots ,2g)$. $w_{\rho}$ is related to the generators
$\alpha_{i}$ of
$H^{1}(M,{\bf{Z}})$ as $w_{\rho}=\sum_{i=1}^{2g}\lambda_{\rho
i}\alpha_{i}$. A basis of holomorphic one-forms on $(M)^{N}$ is given
by
\begin{equation}
w_{\rho k}=1\otimes \cdots \otimes 1\otimes w_{\rho} \otimes 1\otimes
\cdots \otimes 1, \; 1\leq \rho \leq g,\; 1\leq k\leq N       
\end{equation}
with $w_{\rho}$ being in the $k$-th place. Then, a basis $\zeta_{\rho} $ of
holomorphic one-forms on $M_{N}$ is given by the following symmetric
linear combinations
\begin{equation}
\zeta_{\rho} =w_{\rho 1}+\cdots +w_{\rho N},\; 1\leq \rho \leq g.
\end{equation} 
One sees that 
\begin{equation}
\zeta_{\rho}=\sum_{i=1}^{2g}\lambda_{\rho i}\xi_{i}.
\end{equation}
Using the Riemann bilinear relations the
period matrix can be written as $\Lambda
^{t}=(I\; \, \Gamma )$ where $I$ is the $(g\times g)$ unit matrix and
$\Gamma =(\gamma_{jl}),\; (j,l=1,\cdots ,g)$ is a
symmetric matrix with $Im(\Gamma )>0$. 
Notice that under the diffeomorphisms of $M$ the elements
$(\gamma_{jl})$ can change.

Let $v\in H^{2}(M_{N},{\bf{Z}})$ be expressed as
\begin{equation}
v=\frac{1}{2}\sum_{i,j=1}^{2g}q_{ij}\xi_{i}\xi_{j}
\end{equation}
where $Q=(q_{ij})$ is an antisymmetric matrix with integer elements.
Then, expressing $v$ in terms of $\zeta_{\rho}$ one can show that it is of type
(1,1) if the following constraint is satisfied \cite{GH}
\begin{equation}
\Lambda^{t} Q^{-1} \Lambda =0.
\end{equation}
This being a matrix constraint leaves one to freely choose $g^{2}$
elements of $Q$. 
However, for $v$ to be invariant under diffeomorphisms of $M$, the
above equation must be satisfied for arbitrary values of $(\gamma_{jl})$
with $Im(\gamma _{jl})>0$. This can be true only if $Q$ has the
following form
\begin{equation}
Q=D'\left( \begin{array}{cc}
                       0&I \\
                       -I&0

          \end{array}
\right)
\end{equation}
where $I$ is the $(g\times g)$ unit matrix and $D'$ is a constant integer.
Thus, on $M_{N}$ any integral (1,1) form $v$ must
be of the following type
\begin{equation}
v=D'(\sigma_{1}+\cdots +\sigma_{g}).
\end{equation}

\noindent
{\sl (ii) Proof of (3.17)}

Consider the mapping $j: M'\times M'' \rightarrow
M_{N},$ given
by $j(z_{1},z_{2})=(z_{1},\cdots ,z_{1},z_{2},\cdots ,z_{2})$ where
$z_{1}$ occurs $m$ times and $z_{2}$ occurs $(N-m)$ times. $M'$ and
$M''$ are two copies
of $M$. For $z_{2}$ fixed, the mapping
$j$ is an isomorphism onto the submanifold ${\tilde{M}} .$
One obtains
\begin{equation}
j^{\star}(\xi_{i})=m\alpha '_{i}\otimes 1,\; \; j^{\star}(\eta
)=m\beta '\otimes 1.
\end{equation}
Here, $\alpha '_{i}$ and $\beta '$ are, respectively, the generators of
$H^{1}(M', {\bf{Z}})$ and $H^{2}(M',{\bf{Z}})$. Now,
\begin{equation}
\int_{{\tilde{M}} }\eta
=\int_{M'}j^{\star}(\eta)=m\int_{M'}\beta '=m
\end{equation}
and, similarly, since $\sigma_{i}=\xi_{i}\xi_{i+g}$,
\begin{equation}
\int_{{\tilde{M}} }\sigma_{i}=m^{2}, \; 1\leq i \leq g.
\end{equation}
Thus,
\begin{equation}
\int_{{\tilde{M}}}\omega =m(A+2\pi C(g,N)+2\pi mgD(g,N))
\end{equation}
as claimed.

\vspace{.3cm}
{\centerline{\bf{Acknowledgements}}}
\vspace{.25cm}
\noindent
We are grateful to Professor G. Segal for very helpful
discussions. Many thanks
to Dr C. Houghton and Dr H. Merabet for their critical
comments on this manuscript. The work of S.M.N.
was supported by the Overseas Research Council,
the Cambridge Commonwealth Trust and Wolfson College.

This work was partly supported by EPSRC grant GR/K50641, part of the
Applied Nonlinear Mathematics programme.

\end{document}